\documentclass[prd, showpacs,nofootinbib,amsmath,amssymb]{revtex4}
\usepackage{amsfonts}
\usepackage{bm}
\usepackage{verbatim}
\usepackage{graphicx}

\graphicspath{{pics/}}

\usepackage{color}
\definecolor{purple}{rgb}{0.8,0,0.6}

\begin{document}

\title{ Standard Model as the topological material}

 \author{G.E.~Volovik}
\affiliation{Low Temperature Laboratory, School of Science and Technology, Aalto University, Finland}
\affiliation{L.D. Landau Institute for Theoretical Physics, Moscow, Russia}

\author{M.A.~Zubkov}
\affiliation{LE STUDIUM, Loire Valley Institute for Advanced Studies, Tours and Orleans, 45000 Orleans France}
\affiliation{Laboratoire de Mathematiques et de Physique Theorique, Universite de Tours, 37200 Tours, France}
\affiliation{Institute for Theoretical and Experimental Physics,
B. Cheremushkinskaya 25, Moscow, 117259, Russia}


\date{\today}

\begin{abstract}
Study of the Weyl and Dirac topological materials (topological semimetals, insulators,
 superfluids and superconductors) opens the route for the investigation of the topological quantum vacua of relativistic fields.
The symmetric phase of the Standard Model (SM), where both electroweak and chiral symmetry are not broken, represents the topological semimetal. The vacua of the SM (and its extensions) in the phases with broken Electroweak symmetry represent the topological insulators of different types. We discuss in details the topological invariants in both symmetric and broken phases and establish their relation to the stability of vacuum.
\end{abstract}

\maketitle

\section{Introduction}

The massless (gapless) Weyl fermions in symmetric phase of the Standard Model (SM) of fundamental interactions have common topological properties with the Weyl and Dirac fermions in topological semimetals.
 Topological stability of the Weyl node in the spectrum of neutrino was first considered in Ref. \cite{NielsenNinomiya1981}, see also \cite{FrogNielBook}.  Later the topological invariant for the Weyl points  has been expressed in terms of the fermionic Green's function\cite{GrinevichVolovik1988}, and then the topological approach has been extended by  Ho\v{r}ava to the other types of
nodes in the fermionic spectrum, such as Dirac nodal lines and Fermi surfaces \cite{Horava2005};
this topological classification of the possible types of zeroes in the spectrum was based on the $K$-theory.
Topological classification then has been extended to the other topological phases of matter -- the fully gapped states, such as topological insulators, topological superconductors and the phase B of superfluid $^3$He, see \cite{HasanKane2010,Xiao-LiangQi2011,Schnyder2008,Kitaev2009}.

The systems (vacua) with the Weyl points both in condensed matter and in particle physics experience many exotic properties, such as the chiral anomaly. For example, the Adler-Bell-Jackiw equation, which describes the anomalous production of fermions from vacuum \cite{Adler1969,BellJackiw1969,Adler2005} has been  verified in experiments with skyrmions in the chiral superfluid $^3$He-A \cite{BevanNature1997}, see also Ref. \cite{Volovik2003}.
Weyl fermions in semimetals have been considered by Abrikosov and Beneslavskii in 1970
\cite{AbrikosovBeneslavskii1971},
for  the recent reviews on Weyl fermions in semimetals, superconductors and superfluids see  \cite{Soluyanov2015,SchnyderBrydon2015,Mizushima2016,Bansil2016,Yonezawa2016}.

In the topological classification the important role is played by the symmetry of the vacuum.
This also concerns the symmetric phase of SM,  where both electroweak and chiral symmetries are not broken.
In this phase due  to the equal number of the left-handed and the right-handed particles (if the sterile neutrino is included), and due to Lorentz invariance, the total topological charge of the Fermi point situated at ${\bf p}=0$ is zero. Therefore, the topology of the Weyl fermions in the SM is to be supported by symmetry\footnote{It is possible that the Weyl points of the SM originate from the nontrivial topology of the underlying vacuum with Majorana fermions \cite{ZubkovVolovik2013,ZubkovVolovik2014a,ZubkovVolovik2014b}}.
The modification of the momentum space topological invariants associated with various elements of the SM gauge group  has been suggested in Ref. \cite{Volovik2003}.

 In the present paper we consider the complete set of the topological invariants for the Fermi point of the SM fermions. We demonstrate that the generating functional for those invariants possesses the $Z_6$ symmetry, which, relates the elements of the gauge group giving rise to equivalent topological invariants. This is the same $Z_6$ symmetry of the fermionic representations of the SM, which was discussed in \cite{Zubkov:2006zb}. The existence of this $Z_6$ symmetry explains, in particular, why the topological invariant that protects all massless SM fermions may be expressed through either the hypercharge generator or through one of the generators of the $SU(2)_L$ subgroup of the SM gauge group (see Sect. 12.3 of \cite{Volovik2003}).

Next, we discuss the vacua of the SM (and its extensions) in the phases with broken Electroweak symmetry. These phases  represent the topological insulators of different types.
First we consider the conditions at which the Parity breaking interactions may be neglected.
This in partucular requires that the temperature is much smaller than the masses of the corresponding fermions. In this limit the topological classes of the SM vacua are classified according to the topological invariant associated with the matrix of $CT$ symmetry, which protects the number of massive Dirac fermions.
 Notice, that if the interactions would be neglected at all, the vacuum of the SM in the massive phases would be described by the same topology as the   fully gapped superfluid $^3$He-B \cite{Volovik2009}.     In superfluid $^3$He-B the topological invariant is protected by the chiral symmetry of the system. The $CT$ symmetry plays in the  SM the similar role. Presumably, the corresponding topological invariant is relevant for the topological classification of the vacua of the SM at low enough temperatures (pressure, chemical potential etc).

In addition to the topological invariant protected by $CT$ symmetry there exists the topological invariant protected by $T$ symmetry, which is relevant for the consideration of the SM, when the interactions that break $CP$ are taken into account. This invariant becomes important when the emphasis is in the consideration of the Higgs sector of the SM. The topologically nontrivial phase appears, when the Majorana masses of the left - handed neutrinos are present, the number of which is protected by this topological invariant. We demonstrate, that in the noninteracting case of the massive SM Dirac fermions the value of the symmetry protected topological invariant associated with $T$ is equal to zero, $N_{K_T}=0$.
 At the same time in the extensions of the SM with the type II neutrino seesaw the value of the topological invariant $N_{K_T}$ (supported by the $T$-symmetry) is nonzero. Therefore, the phases with and without Majorana masses of the left handed neutrinos cannot be continuously connected and are, indeed, the different phases. However, we obtain, that the type I seesaw is topologically trivial and its vacuum may be transformed without the phase transition to the conventional vacuum of the SM with Dirac neutrino masses.

Depending on the external conditions (temperature, pressure, chemical potentials of various types, etc) the SM and its extensions may exist in various phases. For example, in addition to the ordinary baryonic phase, which is realized at the vanishing temperature, pressure and baryonic chemical potential, in QCD there exist various other phases: several color supercondicting phases, the quark - gluon plasma phase,  etc  \cite{CFL0,CFL,m1,m1m8}. The Weinberg - Salam model is typically considered in the two phases: the symmetric high temperature phase with the restored chiral symmetry and the broken low temperature phase with the spontaneously broken $SU(2)\otimes U(1)$ symmetry. The complete SM (containing QCD and the Weinberg - Salam model) may possess new phases, which were not considered yet, at certain external conditions. Various extensions of the SM like the models with Majorana masses of neutrinos, the models with several Higgs bosons, the models with composite Higgs bosons may also exist in several exotic phases, which were not considered so far.   The momentum space topological invariants discussed in the present paper (or their extensions) may be applied to the consideration of the phase transitions between the phases of the SM (and its extensions) mentioned above. For the previous consideration of the topologically nontrivial vacua in relativistic quantum field theories based on the topological invariants in momentum space see, for example,
\cite{So1985,IshikawaMatsuyama1986,Kaplan1992,Golterman1993,Volovik2003,Horava2005,Creutz2008,Kaplan2011,Creutz2011,ZubkovVolovik2012,Zubkov2012}.

The paper is organized as follows. In Sect. \ref{SymmetricPhase} we consider the symmetric phase of the SM with unbroken chiral and Electroweak symmetries as the phase of the topological semimetal. The complete set of the topological invariants protecting the Weyl points is defined, the $Z_6$ symmetry of the corresponding generating functional is established.  In Sect. \ref{Parity} and Sect. \ref{Tsymmetry} we discuss the SM at low temperatures, which are smaller than the mass of the lightest fermion. In Sect. \ref{Parity} we discuss the situation at the sufficiently small values of pressure and chemical potentials, so that the parity breaking interactions are to be neglected in the considerations of the questions of the stability of vacuum.  In Sect. \ref{Tsymmetry} we discuss the topological invariant of the SM and its extensions that remains at work if the parity breaking interactions are taken into account while the CP breaking is neglected. In particular, it is demonstrated that the vacuum with Dirac fermions is topologically trivial (with respect to the invariant protected by time reversal symmetry). At the same time, the vacuum with Majorana masses may be topologically nontrivial.

\section{Standard Model in the symmetric phase as the topological semimetal}
\label{SymmetricPhase}

\subsection{Topological invariant for massless fermions}


In its gapless (massless) phase the Standard Model belongs to the class of the  3+1 dimensional vacua, which  are characterized by the Weyl points in momentum space.  The Weyl point is characterized by the momentum space topological invariant $N_3$, which protects the masslessness of the fermionic spectrum.  The topological invariant for the isolated Weyl point is expressed as the integral
of  the 3-form in terms of the two - point Green's function $\cal G$ determined in the 4 dimensional momentum-frequency space {\cite{GrinevichVolovik1988,Volovik2003}:
\begin{equation}
N_3 = {\bf tr} ~{\cal N}~~,~~{\cal N}  =
\frac{1}{{24\pi^2}}e_{\mu\nu\lambda\gamma}~
 ~\int_{\sigma}~  dS^{\gamma}
~ {\cal G}\partial_{p_\mu} {\cal G}^{-1}
{\cal G}\partial_{p_\nu} {\cal G}^{-1} {\cal G}\partial_{p_\lambda}  {\cal
G}^{-1}~.
\label{TopInvariantMatrix}
\end{equation}
The Green's function is $n\times n$ matrix.  For a single species of Weyl fermions  one has $n=2$, and  the Green's function  is expressed in terms of the Pauli matrices.
For general topological condensed matter system the $n\times n$ matrix contains Pauli matrices for spin and for Bogoliubov spin, and also the crystal band indices of fermions. In particle physics, the $n\times n$ matrix includes Weyl or Dirac matrices and  indices of different fermionic species  (quarks and  leptons of different generations). In SM with 16 species in one generation one has $n=32g$ , where $g$ is the number of generations. If expressed in terms of Majorana fermions, the matrix has $n=64g$.  The definition of the Green function in terms of the functional integral over the fields is given in Appendix A.
The integral in  Eq. (\ref{TopInvariantMatrix}) is over the $S^3$ surface
$\sigma$ embracing the point in the 4D space ${\bf p}=0, p_4=0$, where $p_4$¸ is the  frequency
along the imaginary axis; ${\bf tr}$ is the trace over the fermionic indices.
For  a single species of right handed Weyl fermions
one has $N_3=1$, and $N_3=-1$ for the left handed Weyl fermions.

It is worth mentioning that the symmetric phase of the SM appears at the finite temperatures,
while the topological classification and topological invariants are formally applicable only to the ground state (vacuum) of the system. Actually the consideration is valid if the temperature $T$ is much smaller than the characteristic high energy scale of the system $T\ll T_{\rm uv}$. Here $T_{\rm uv}$ is the scale, at which the Standard Model of fundamental interactions already does not work, and the new fields and interactions appear. In this limit all the properties of the systems related to topology, such as chiral anomaly, are determined by these topological invariants.
Thus in spite that the topological invariant $N_3$ is defined typically for the zero temperature, in the SM this invariant appears to be well - defined at the temperatures above the Electroweak transition if those temperatures are smaller than the scale of the ultraviolet completion of the SM, which is at least one order of magnitude higher than the Electroweak scale $\sim 100$ GeV.

 If the sterile right handed neutrinos are present in the Standard Model, the number of the left and the right handed fermions is equal, $n_{\rm left}= n_{\rm right}= 8g$, where $g$ is the number of generations. This is required, for example, if we assume that the lattice regularization is used, where the number of the left - handed and the right - handed fermions are equal due to the Nielsen - Ninomiya theorem.
 Then the trace in
 Eq.(\ref{TopInvariantMatrix}) over all the fermionic species gives the trivial value for the
 topological invariant, $N_3=   n_{\rm right}- n_{\rm left}=0$. Nevertheless, the vacuum  of the Standard Model is topologically nontrivial, because its topology is supported by the symmetry of the SM in the symmetric phase.
 The $SU(3)\otimes SU(2) \otimes U(1)$ symmetry allows to introduce the generating function of topological invariants,
which contains the powers of the hypercharge $Y$, the generators of $SU(2)_L$ and $SU(3)_c$:
\begin{equation}
N(\theta_Y, \theta^a_W, \theta^i_c)= {\bf tr} \, \left[ e^{i\theta^a_W {\cal W}_a } e^{i\theta_Y
{\cal Y}  } e^{i\theta^i_c {\cal C}_i } {\cal N}
\right] \,
\label{GeneratingFunctionWYC}
\end{equation}
Here ${\cal W}_a$, $a=1,2,3$ are the generators of $SU(2)_L$ while  ${\cal C}_i$, $i=1,...,8$ are the generators of $SU(3)_c$.

\subsection{The $Z_6$ symmetry of the fermionic representations in the Standard Model.}

Notice, that Eq. (\ref{GeneratingFunctionWYC}) obeys the $Z_6$ invariance (see also \cite{Bakker:2003gg}):
\begin{eqnarray}
\theta_Y  \rightarrow \theta_Y + 2 \pi N, \quad e^{i\theta^a_W {\cal W}_a }  \rightarrow e^{i\theta^a_W {\cal W}_a }  \times e^{i \pi N} , \quad  e^{i\theta^i_c {\cal C}_i } \rightarrow  e^{i\theta^i_c {\cal C}_i } \times e^{\frac{2\pi i}{3}N} \label{Z6}
\end{eqnarray}
where $N$ is integer. This invariance might actually mean, that the gauge group of the SM is $SU(3)\otimes SU(2) \otimes U(1)/Z_6$ rather than $SU(3)\otimes SU(2) \otimes U(1)$ \cite{Zubkov:2006zb}.
The given $Z_6$ symmetry follows from the  assignment of the hypercharges, weak charges and electric charges $Q=Y+W$   of the fermions
given by
\begin{equation}
\begin{array}{cccc}  {\rm Fermion} &W&Y&Q\\ u_L(3)&+{1\over 2}&{1\over 6}&{2\over 3}\\ u_R(3)&0
&{2\over 3}&{2\over 3}\\ d_L(3) & -{1\over 2}&{1\over 6}&-{1\over 3}\\ d_R(3)
&0   &-{1\over 3}&-{1\over 3}\cr \nu_L &+{1\over 2}   &-{1\over 2}&0\\ \nu_R
&0   &0&0\cr e_L &-{1\over 2}  &-{1\over 2}&-1\cr e_R &0   &-1&-1\cr \end{array}
\label{SU42}
\end{equation}
According to this table in the Standard Model including  strong interactions
the group $U(1)\times SU(2)\times SU(3)$  has the global $Z_6=Z_2\times
Z_3$-subgroup of elements which act on the SM fermions as identity element (see
eqs. (61)-(64) in \cite{FrogNielBook} and references
\cite{Bakker:2003gg,Zubkov:2006zb}). This group consists of the following elements
$g^k$:
\begin{equation} g^k=\left[ e^{i 2\pi {\cal C}_8}e^{2\pi  i {\cal
Y}  } e^{2\pi  i {\cal W}_3  }\right]^k   ~~,~~ k=1,...,6 ~. \label{Z3}
\end{equation}
where ${\cal C}_8$ will be specified below in Eq. (\ref{C8}).
Notice, that the $Z_6$ symmetry of the fermionic representations of the SM takes place in any phases, not only in the symmetric phase. Its elements of Eq. (\ref{Z3}) being applied to any fermion of the SM give $1$. In the other words, all SM fermions represent the eigenvectors of the elements of $Z_6$ corresponding to the eigenvalues equal to unity.

\subsection{Maximal number of Dirac massless fermions protected by the topological invariants}

The generators ${\cal W}_a $, ${\cal C}_i$, ${\cal  Y}$   commute with the fermion Green functions taken in Landau gauge. Therefore, there is the following global $SU(2)\otimes SU(3)$ invariance:
\begin{eqnarray}
e^{i\theta^a_W {\cal W}_a }  \rightarrow U e^{i\theta^a_W {\cal W}_a }  U^+, \quad  e^{i\theta^i_c {\cal C}_i } \rightarrow  \Gamma e^{i\theta^i_c {\cal C}_i }\Gamma^+
\end{eqnarray}
where $U\in SU(2)$ while $\Gamma \in SU(3)$. As a result we can represent Eq. (\ref{GeneratingFunctionWYC}) in the form:
\begin{equation}
N(\theta_Y, \theta_W, \theta_c, \theta^\prime_c)= {\bf tr} \, \left[ e^{i\theta_W {\cal W}_3 } e^{i\theta_Y
{\cal Y}  } e^{i\theta_c {\cal C}_8 +  i\theta^\prime_c {\cal C}_3} {\cal N}
\right] \,
\label{GeneratingFunctionWYC2}
\end{equation}
where for the left - handed doublets of fermions
\begin{equation}
 {\cal W}_3  = \frac{1}{2}\left(\begin{array}{cc} 1 & 0 \\0 & -1\end{array}\right)
\end{equation}
while for the colored quarks
\begin{equation}
 {\cal C}_8  = \frac{1}{3}\left(\begin{array}{ccc} 1 & 0 & 0 \\0 & 1 & 0 \\ 0 & 0 & -2\end{array}\right), \quad {\cal C}_3  = \frac{1}{2}\left(\begin{array}{ccc} 1 & 0 & 0 \\0 & -1 & 0 \\ 0 & 0 & 0\end{array}\right)\label{C8}
\end{equation}

The direct calculation gives
\begin{equation}
N(\theta_Y, \theta_W,\theta_c,\theta_c^\prime)=
2g\left(\cos {\theta_Y\over 2} -\cos {\theta_W\over 2}\right) \left(
e^{i
\theta_{Y}/6} (e^{\frac{i \theta_c}{3}+i\frac{\theta_c^\prime}{2}}+e^{\frac{i \theta_c}{3}-i\frac{\theta_c^\prime}{2}}+e^{-\frac{2i \theta_c}{3}})+   e^{-i
\theta_{Y}/2}\right)
\label{GeneratingFunctionWY}
\end{equation}
The particular case of this expression with $\theta_c = \theta_c^\prime=0$ was considered, for example, in \cite{Volovik2000,Volovik2003}. On the level of the angles $\theta_Y, \theta_W,\theta_c,\theta_c^\prime$ the $Z_6$ symmetry has the form:
\begin{equation}
\theta_Y \rightarrow \theta_Y+2 \pi N,\quad  \theta_W \rightarrow \theta_W+2\pi N,\quad \theta_c\to\theta_c + 2\pi N,\quad \theta_c^\prime\to \theta_c^\prime \label{Z62}
\end{equation}
(Notice that $\theta_W$ is defined modulo $4\pi$, $\theta_c$ is defined modulo $6\pi$, while $\theta_Y$ is defined modulo $12\pi$.)
The generating function is robust to the deformations of the Green's function, if those deformations obey the SM symmetry.

The choice of parameters ($\theta_Y=0$,  $\theta_W=2\pi$, {$\theta_c=\theta^\prime_c=0$})
and any other choice related to it by the $Z_6$ transformation of Eq. (\ref{Z62})
gives the maximally possible value of the generating function:
\begin{equation}
 N_{\rm max}=   N(\theta_Y=0, \theta_W=2\pi,\theta_c=0,\theta_c^\prime=0)  = 16g \,.
\label{16g}
\end{equation}
This value guarantees that all 16$g$ fermions of the Standard Models are massless in the symmetric phase.
Those maximal values (\ref{16g}) are formed by the discrete subgroup of the Standard Model symmetry group \cite{Volovik2010} (that is related by the $Z_6$ transformation to the centers of $SU(3)_c$ and $SU(2)_L$):
\begin{equation}
 N_{\rm max}=    {\bf tr} ~\left[K_Y{\cal N}\right]= 16g \,,
\label{max}
\end{equation}
where we may take
 \begin{equation}
K_Y=  e^{i [2\pi (N+1)]_{{\rm mod}\,4\pi} {\cal W}_3 } e^{i [2\pi N]_{{\rm mod}\,12\pi}
{\cal Y}  } e^{i [2\pi N]_{{\rm mod}\,6\pi} {\cal C}_8}\label{KY}
\end{equation} with any integer $N$. In particular, for $N=3$ we get $K_Y =e^{6\pi i Y}$ while for $N=0$ we have $K_Y= e^{i  2\pi  {\cal W}_3 } $. Various operators in Eq. (\ref{KY}) are related by the $Z_6$ transformation.


\section{SM at low temperatures as the topological insulator with C, P, and T symmetries.}
\label{Parity}

The topology of the SM vacuum in the massive phase looks similar to that of the ground state of superfluid $^3$He-B, which in the noninteracting case is described by the integer valued topological invariant \cite{Volovik2009}
\begin{equation}
N_K =
\frac{1}{{24\pi^2}}e_{\mu\nu\lambda} \, {\bf tr} \, \left[ K
 ~\int  d^3p
~ {\cal H}^{-1}\partial_{p_\mu} {\cal H} \,
{\cal H}^{-1}\partial_{p_\nu} {\cal H} \, {\cal H}^{-1}\partial_{p_\lambda}  {\cal
H} \right]\,.
\label{TopInvariantMassive}
\end{equation}
Here ${\cal H}({\bf p})={\cal G}^{-1}(p_4=0)$  (while $\cal G$ is the Green function); the integral is over 3-momentum space;
and $K$ is the proper symmetry operation (it should either commute of anti - commute with $\cal H$).  For the superfluid $^3$He-B one has $N_K=2$ for $K= \tau_2$ (the combination of time reversal  and particle-hole symmetries \cite{Volovik2009}). The larger values of this invariant may be obtained by the extension of the model of $^3$He-B to the multi - component fermionic models\cite{WangYang2016}.

The expression for Eq. (\ref{TopInvariantMassive}) is formally defined at zero temperature, $T=0$. However the effects of the nontrivial topology on the physical properties of the systems
can be measured at finite temperatures. For some effects the temperatures must be much smaller than the masses of the fermions existing in the given system, $T\ll m$, while for the others the limit
$1/ \tau(T) \ll m$  is enough, where $\tau(T)$ is the characteristic relaxation time.   In the latter case
it is not excluded that the topological invariant can be applicable even for $T >m$.  Possibly, the definition of topological invariant may be extended even further, but we do not discuss here this possibility.

The smallest Dirac mass in the SM is the mass of electron. The leading term in the temperature corrections to the corresponding self energy is the one loop expression proportional to the fine structure constant $\alpha$. This term gives rise to the shift of the dispersion of the quasiparticles by the amount of the order of $e T$ \cite{Ballac}.  Therefore, the requirement $e T \ll m_e$ gives $ T \ll 1$ MeV. Even more strong restriction comes from the Neutrino sector, where the thermal contribution to mass may be roughly given by expression $g T$ \cite{Ghiglieri:2016xye}, where $g$ is the $SU(2)$ or $U(1)_Y$ coupling constant. Assuming that the neutrino mass is about $1$ eV, we come to the restriction $T\ll 1$ eV. This condition is satisfied, for example, by the present state of the Unverse with the temperature of the order of $10^{-4}$ eV.

The question arises, whether the vacuum of the SM in  the massive phase is topologically trivial or not. If yes, what is the corresponding matrix $K$ for the Standard Model and what is the effect of interactions.  The situation here is completely unclear. First of all, we may consider the approximation to the SM, in which the exchange by the W and Z bosons as well as the Higgs boson are neglected. Roughly, this may correspond to the description of processes at the energies much smaller than the electroweak scale $\sim 100 $ GeV.
Then in the Standard Model
at zero temperature $T=0$ or at nonzero $T$ with the proper restrictions such as  $T\ll m$, where $m$ is the smallest fermion mass in the phase with the spontaneously broken electroweak symmetry, the Green's function has the form
 \begin{equation}
{\cal G}(p)=  Z(-p^2) \frac{1}{\gamma^\mu p_\mu- M(-p^2)}\,,
\label{eq:S(p)}
\end{equation}
where $p^2=-{\bf p}^2+\omega^2$, while $Z$ and $M$ are matrices. The fermion mass matrix $m$ is given by the solution of equation
  $$M(-m^2) = m$$
while Dirac matrices may be chosen according to Sec. 5.4 in Ref. \cite{WeinbergQTF}:
  \begin{equation}
  \gamma^0=\tau_1~~,~~ {\mbox{\boldmath$\gamma$}}= i \tau_2{\mbox{\boldmath$\sigma$}}~~,~~
  \gamma_5=-i\gamma^0\gamma^1\gamma^2\gamma^3=\tau_3
 \,.
\label{eq:DiracMatrices}
\end{equation}
This approximation is reasonable due to the smallness of the fine structure constant and large enough masses of W, Z, and the SM Higgs boson. Here it is the matrix  $K=-i\gamma^5\gamma^0=\tau_2$, which commutes with the Green's function at $\omega=0$. This is the matrix of the combination of CPT and P transformations\footnote{The unessential phase factor of this symmetry matrix is chosen in such a way, that the expression of Eq. (\ref{TopInvariantMassive}) gives the real value for the case of the non - interacting Dirac fermions.}, that is at the same time the combination of C and T.
  As a result Eq. (\ref{TopInvariantMassive}) determines the topological invariant. It may be calculated for the simplest system connected with the given one by a continuous transformation. Assuming that such a connection exists with the system of noninteracting massive Dirac fermions (that represent the constituents of the SM), we obtain
 \begin{equation}
N_K=8g \,,
\label{eq:8g}
\end{equation}
where $g$ is the number of generations of the SM fermions\footnote{For the system of the non-intreracting Dirac fermions with masses $M_a$ the given topological invariant is given by
$N_K =\sum_a {\rm sign} M_a$.}. In notations of Ref.\cite{Metlitski2014,Ryu2015} the invariant is $\nu=N_K/2$.

At the present moment the role of the interactions between the fermions is not completely clear. For example, even at zero temperature the strong $SU(3)$ interactions give rise to the transition between the system of the noninteracting quarks and the confining QCD. This may (but also may not) give rise to the value of the topological invariant associated with the CT symmetry that differs from the value calculated above. The answer depends on the possibility to transform continuously the two point fermion Green function for the noninteracting massive fermions to the two point Green function of QCD with the strong interactions taken into account. Various approximations to QCD may give different answers to this question. For example, the Nambu - Jona - Lasinio (NJL) approximation allows to connect continuously the interacting and non - interacting Green functions. The spectrum of the lightest resonances is described by the NJL model reasonably well. This allows to suppose, that in the low energy effective theory Eq. (\ref{eq:8g}) gives the correct answer for the hadronic phase of the SM.

 Formally the parity breaking interactions destroy the consideration of the  topological  stability based on the invariant $N_{K_{CT}}$. In practice, this invariant remains operative because of the smallness of the corrections. But the topological classification group may be reduced from $Z$ to the smaller group. For example, the Electroweak $SU(2)$ interactions assume, that the vacua with the values of  $N_{K_{CT}}$ of opposite signs (that correspond to the opposite values of masses of all fermions) represent the same physical vacuum. This occurs because during the Electroweak symmetry breaking the opposite values of masses appear as different versions of Unitary gauge. In the complete theory with the $SU(2)$ interactions taken into account those states are related by global gauge transformation, and therefore, not only connected continuously but represent the same physical vacuum. This reduces the topological classification to $Z/Z_2$. The reduction may be more significant if the SM appears as a low energy approximation to a certain theory with the larger gauge group.  For the recent discussion of the similar modification of topological stability pattern in topological superconductors due to interactions see
  \cite{Kitaev,Metlitski2014,Ryu2015,Tachikawa2016,Witten2015,Yonekura2016,Ryu2016,Wang2014,Kapustin2015}.
In practice the reduction of the topological classification means that various defects lose their topological stability. For example, let us consider the QCD sector of the SM with the two quarks (u and d). Let us also neglect the current masses of quarks,  There is the chiral $SU(2)$ symmetry which is broken spontaneously in the hadronic phase. As a result the constituent quark masses appear. The positive and negative values of masses may appear in this way. One may naively suppose, that this should lead to the formation of the topologically stable domain walls separating the regions with the opposite values of  the constituent masses. But this is actually not so. The opposite values of masses appear as the arbitrary choice of the sign of the condensate. Those choices are related by the element of the global chiral symmetry $SU(2)$.
We may consider the version of the theory with the $2\times 2$ complex - valued condensate field, and in this theory the state with positive mass is connected continuously by the symmetry transformation with the state with negative mass. (At the intermediate states the mass in undefined.) This means the reduction $Z\to Z/Z_2$ of the symmetry classification and this means that the topologically stable domain walls in the Hadronic phase of QCD do not exist.
 In practice if such domain walls appear dynamically in the form of bubbles, then they decay with the emission of the $SU(2)\times U(1)$ gauge bosons.

\section{The topological invariant protected by T  symmetry}
\label{Tsymmetry}

\subsection{The version of the Standard Model with Majorana masses of left - handed neutrinos}
\label{MassiveNu}

In Section \ref{Parity} we considered the approximation to the SM when Parity remains unbroken.  The question of the stability of vacuum was related to the topological invariant protected by $CT$. Interactions with the SM Higgs boson and with the W and Z bosons destroy the vacuum stability criteria based on the consideration of this invariant. At least, the interactions in the Higgs sector are strong. Although we may neglect this effect in some approximation, this is necessary to consider the other topological invariants. In order to consider such invariants we use the representation of the SM in terms of the Nambu - Gorkov spinors. This  allows to treat the particle - antiparticle transformation as matrix. We assume in the present section that weak CP breaking interactions do not affect the stability of vacuum. Therefore, we will use the topological invariant protected by $T$ symmetry.

In this subsection we consider the version of the Standard Model with the left - handed massive neutrinos. First of all, let us discuss
the situation, when the right - handed neutrinos remain massless, the Dirac masses of neutrinos are absent, and only the observed left - handed neutrinos are massive. For example, the type II neutrino seesaw may lead to such pattern. Then the following mass term appears \begin{equation}
L_\nu =-M_{} \nu^{A}_L \nu_L^B \epsilon_{AB}+ (h.c.)
\end{equation}
(summation over the generations is implied).

Let us consider the situation, when the interacting system may be deformed continuously to the system without interactions in the lepton sector and with Majorana masses of the left - handed neutrinos.  Let us introduce the conventional definition of the Nambu - Gorkov spinor:  ${\cal N}_L = (\nu^{c}_L, \nu_L)^T$ (where $\nu_L^{c} = i\sigma^2 \bar{\nu}_L$). In terms of this spinor the lagrangian for one massive non - interacting left - handed neutrino may be written as follows:
\begin{equation}
L_L = \bar{\cal N}_L( \gamma^\mu p_\mu  + M) {\cal N}_L
\end{equation}
where
\begin{equation}
\bar{\cal N} = {\cal N}^Ti \gamma^2\gamma^0
\end{equation}
The time reversal transformation reads:
\begin{equation}
{\cal N}_L \rightarrow \gamma^0 \gamma^5 {\cal N}_L
\end{equation}
we choose the unessential phase factor in such a way, that  in this representation the matrix of the time reversal transformation is given by
\begin{equation}
{\bf K}_T =-i \gamma^0\gamma^5
\end{equation}
The corresponding  topological invariant receives the form:
\begin{equation}
N_{K_T} = {e_{ijk}\over{48\pi^2}} ~
{\bf tr}\left[  \int_{\omega=0}   d^3p ~{\bf K}_T
~G\partial_{p_i} G^{-1}
G\partial_{p_j} G^{-1} G\partial_{p_k} G^{-1}\right]\,.
\label{3DTopInvariant_tau_24e}
\end{equation}
with
\begin{equation}
 G^{-1} =  \gamma^\mu p_\mu  + M
\end{equation}
We obtain:
\begin{eqnarray}
N_{K_T} &=& i{e_{ijk}\over{48\pi^2}} ~
{\bf tr}\left[  \int_{\omega=0}   \frac{d^3p}{(\vec{p}^2+M^2)^2} ~\gamma^0\gamma^5( + M  )\gamma^i\gamma^j\gamma^k
~ \right]\nonumber\\
&=&  \frac{1}{8}\Big(2\times 4\times \frac{1}{2} \Big) = 1/2
\end{eqnarray}
If we have $g$ generations of the left - handed neutrinos, the result is to be multiplied by $g$. As we will see below the quark sector with Dirac masses does not give the nonzero contribution to $N_{K_T}$. Therefore, assuming that the SM vacuum in the given phase is connected continuously with the vacuum of the version of the SM without interactions between leptons, we obtain the overall value of the invariant
\begin{eqnarray}
N_{K_T} &=& g/2\label{lg}
\end{eqnarray}
 We suppose, that this property takes  place for the SM at vanishing temperature, pressure and chemical potentials.

\subsection{The version of the Standard Model with Type I  neutrino seesaw.}

Let us remind the basics of the classical type I seesaw \cite{Hernandez:2010mi}. In the basis ${\cal N}_{LR} = (\nu^{c}_L, \nu_R)^T$ (where $\nu_L^{c} = i\sigma^2 \bar{\nu}_L$) there is the mass matrix
\begin{equation}
{\bf M}_{\nu} = \left(\begin{array}{cc} 0 & m_{}\\
m_{} &   M_{} \end{array}\right)\label{MASSSEESAW}
\end{equation}
The overall mass term is $\frac{1}{2}{\cal N}_{LR} {\bf M}_{\nu} {\cal N}_{LR} + (h.c.)$, where $(h.c.)$ means hermitian conjugation that implies $\nu_{R,L} \rightarrow \bar{\nu}_{R,L}$ and vice versa. The product of the two - component spinors is defined as
$$ {\cal N}_{LR}{\cal N}_{LR} \equiv {\cal N}_{LR}^A{\cal N}_{LR}^B\epsilon_{AB}$$
For simplicity we assume, that $g$ Dirac masses $m_{}$ are equal to each other and $g$  Majorana masses $M_{}$ are also equal.
 The diagonalization gives $g$ heavy neutrinos with Majorana masses $M_{heavy}  \approx + M_{}$ and $g$  light neutrinos with Majorana masses
 \begin{equation}
 -M_{light} \approx  - m_{} \frac{m_{}}{ M_{}}
\end{equation}
Notice, that the signs of $M_{heavy}$ and $-M_{light}$ are opposite. If we rewrite the mass term through the spinor $\nu_L$ rather than $\nu_L^c$ , then the sign of the mass becomes positive because
\begin{eqnarray}
L_\nu &=& - M_{light} \Big[\nu^c_L\Big]^T (i\sigma^2) \nu_L^c  -   M_{light} \Big[\bar{\nu}^c_L\Big]^T (-i\sigma^2) \bar{\nu}_L^c \nonumber\\ & = &  M_{light} \Big[\bar{\nu}_L\Big]^T (-i\sigma^2) \bar{\nu}_L  +   M_{light} \Big[{\nu}_L\Big]^T( i\sigma^2) {\nu}_L
\end{eqnarray}
 This results in the trivial value of the corresponding momentum space topological invariant (see below).
The assumptions that the mass of the right - handed neutrinos $M_{}$ is not smaller, than $1$ TeV, and that the Dirac neutrino mass $m_{}$ is of the order of the electron mass $m_e$ allow us to estimate
$ M_{light}  \le  0.25$ \, eV.

The topological invariant for the left handed neutrino was calculated above and is given by Eq. (\ref{lg}). Now let us consider the right - handed neutrino. We define
 ${\cal N}_R = (\nu_R, \nu^c_R)^T$ (where $\nu_R^{c} = i\sigma^2 \bar{\nu}_R$). In terms of this spinor the lagrangian for the massive right - handed neutrino may be written as:
\begin{equation}
L_R = \bar{\cal N}_R( \gamma^\mu p_\mu  + M) {\cal N}_R
\end{equation}
Now the time reversal transformation reads:
\begin{equation}
{\cal N}_R \rightarrow - \gamma^0 \gamma^5 {\cal N}_R
\end{equation}
which means that in this representation
\begin{equation}
{\bf K}_T = i \gamma^0\gamma^5
\end{equation}
The  topological invariant is given by the same expression of Eq. (\ref{3DTopInvariant_tau_24e}).
It gives:
\begin{eqnarray}
N_{K_T} &=& -ig{e_{ijk}\over{48\pi^2}} ~
{\bf tr}\left[  \int_{\omega=0}   \frac{d^3p}{(\vec{p}^2+M^2)^2} ~\gamma^0\gamma^5( + M  )\gamma^i\gamma^j\gamma^k
~ \right] =- g/2
\end{eqnarray}
where $g$ is the number of generations.
One can see, that the system with the equal number of the left - handed and the right - handed neutrinos with the Majorana masses of the same sign has the vanishing value of topological invariant.

\subsection{The version of the SM with Dirac masses of neutrinos}

\label{Nambu}

 In the present subsection we discuss the case, when Majorana masses are absent, and follow the alternative definition of the Nambu - Gorkov spinors introduced in \cite{Z2014}. In  Appendix B  we represent the corresponding definition.
In terms of the corresponding Green functions the topological invariant for the SM may be written as
\begin{equation}
N_K = {e_{ijk}\over{48\pi^2}} ~
{\bf tr}\left[  \int_{\omega=0}   d^3p ~{\bf K}
~G\partial_{p_i} G^{-1}
G\partial_{p_j} G^{-1} G\partial_{p_k} G^{-1}\right]\,.
\label{3DTopInvariant_tau}
\end{equation}
where $\bf K$ is the appropriate symmetry of the system while $G$ is the Green function being the vacuum average of the product of spinors $\Psi$ of Eq. (\ref{PSI}) of Appendix. For the definition of the two sets of gamma - matrices $\gamma^\mu$ and $\Gamma^a$ also see Appendix.
 First of all let us consider again the case of the P - invariant approximation to the SM, in which case we may use matrix of CT $${\bf K}={\bf K}_{CT} =
i \Gamma^4  \Gamma^5 \gamma^0 $$
 then if the Green function of the system may be adiabatically deformed to that of the system with free massive Dirac fermions with the same mass $M$, then
 we substitute into Eq. (\ref{3DTopInvariant_tau}) the Green function of the form:
  \begin{eqnarray}
G^{-1}(p)&=&  i \gamma^2 \gamma^5\gamma^0 \Gamma^4 \Gamma^2 \Gamma^5 \Big[\gamma^\mu p_\mu  + M\gamma^5 \Gamma^5 \Gamma^4  \Big]
\label{eq:S(p)20}
\end{eqnarray}
 This gives
\begin{equation}
N_{K_{CT}} = i g(N_c+1){e_{ijk}\over{48\pi^2}} ~
{\bf tr}\left[  \int_{\omega=0}   \frac{d^3p}{(\vec{p}^2+M^2)^2} ~\Gamma^4\Gamma^5\gamma^0(\vec{\gamma}\vec{p} - M_0 \Gamma^4 \Gamma^5 \gamma^5 )\gamma^i\gamma^j\gamma^k
~\right] =2 g(N_c+1)
\label{3DTopInvariant_tau_248}
\end{equation} (where $N_c=3$ is the number of colors) in accordance with the calculation of Sect. \ref{Parity}.


If we take into account interactions that break parity, but neglect the complexness of the elements of fermion mixing matrix, then  the SM is CP - invariant. In this situation the topological invariant may be composed using matrix
$$ {\bf K}_T = -i
\Gamma^4 \Gamma^2 \Gamma^5 \gamma^0 $$
Let us again suppose, that two point fermion Green function of the SM is connected continuously with that of the non - interacting theory with all Dirac masses of the fermions equal to each other. In this case our topological invariant has the form
\begin{equation}
N_{K_T} = -ig(N_c+1){e_{ijk}\over{48\pi^2}} ~
{\bf tr}\left[  \int_{\omega=0}   \frac{d^3p}{(\vec{p}^2+M^2)^2} ~\Gamma^4\Gamma^2\Gamma^5\gamma^0(\vec{\gamma}\vec{p} -M_0 \Gamma^4 \Gamma^5 \gamma^5)\gamma^i\gamma^j\gamma^k
~\right]\sim {\bf tr} \, \Gamma^2 \equiv 0
\label{3DTopInvariant_tau_248}
\end{equation}
Thus, one can see, that the parity breaking interactions make the SM vacuum with Dirac masses of all fermions trivial.

\section{Conclusions}

In the present paper we look at various phases of the Standard Model of fundamental interactions (or its extensions) as at the systems similar to the topological materials. The symmetric phase (the temperature is above the temperature of Electroweak transition) represents the topological semimetal with the Fermi point, which is protected by the topological invariants  generated by the functional Eq. (\ref{GeneratingFunctionWYC2}). This functional depends on the angles $\theta_Y,\theta_W^a, \theta^i_c$. It has been demonstrated that the $Z_6$ symmetry of the fermionic representations of the SM \cite{Bakker:2003gg} manifests itself as the symmetry of this functional under the corresponding transformation of the mentioned angles Eq. (\ref{Z6}). Due to the $Z_6$ symmetry the maximal value of the topological invariant protecting the Fermi point may be constructed either of the hypercharge or of the generator ${\cal W}_3$ of $SU(2)_L$. It is given by   Eq. (\ref{KY}). This maximal value is equal to the number of massless SM fermions.  This is peculiar, that although the formal definition of these topological invariants was given at zero temperature, this is the phase with high temperature, where they may be applied. This occurs because in this phase the mass scale disappears, and the relevant scale parameter is given by temperature itself. It should be compared to the scale $\Lambda$, at which the Standard Model transfers to its ultraviolet completion. We suppose, that the considered here topological invariants protecting the Fermi point remain at work at least as long as $T \ll \Lambda$. Notice, that such a scale may not be smaller than $1$ TeV.

At the temperatures below the Electroweak phase transition the SM (and its extensions) may exist in several phases, where the fermions are massive. Those phases resemble various phases of topological insulators and are characterized by the corresponding topological invariants in momentum space. At small enough temperature, pressure and chemical potentials the parity breaking interactions may be neglected, and the vacuum is topologically nontrivial being  protected by the topological invariant of Eq. (\ref{3DTopInvariant_tau}) with  $\bf  K$ given by the matrix of $CT$ transformation. Our definition of this invariant remains valid at least up to the temperatures much smaller than the neutrino masses, which are assumed to be of the order of $eV$. Although the strong interactions of the SM provide the transition between the system of noninteracting quarks and the quarks confined by the quark - gluon strings, we may suppose, that the two - point quark Green function is connected continuously with that of the non - interacting quarks. At least, this occurs in the NJL approximation to the QCD \cite{NJLQCD}, which describes the light mesons reasonably well. See also the recent lattice data on quark propagator in \cite{Oliveira:2016muq} and references therein, which indicate that the functions $Z(-p^2)$ and $M(-p^2)$  in Eq. (\ref{eq:S(p)}) tend to finite nonzero values\footnote{Though, one cannot exclude considering these data that the quark function $Z(-p^2)$ tends to zero at $p\to 0$, in which case the singularity is encountered in the expression for the topological invariants. This is a difficult question how to regularize such singularities if they do appear, and we omit this question in the present paper.} at $p\rightarrow 0$. Assuming the possibility of such continuous transformation, we come to the value of the topological invariant that protects the number of massive Dirac fermions.

 If we take into account the parity broken interactions, then the topological classification based on the invariant protected by $CT$ is reduced at least to $Z/Z_2$. The further reduction is possible if BSM unified model has the appropriate extended gauge symmetry, We deal with the topological insulator with $T$ symmetry if we neglect weak CP breaking interactions originated from the imaginary parts of the elements of quark mixing matrix.  Therefore, the stability of vacuum is protected also by the topological invariant of Eq. (\ref{3DTopInvariant_tau}) with  $\bf  K$ given by the matrix of the time reversal transformation. This expression remains operative at least for the temperatures $T \ll 1$ eV. We do not exclude, that there exists the extension of its definition to the essentially higher temperatures. But this is out of the scope of the present paper. It appears, that the value of this invariant for the SM with massive Dirac neutrinos is zero.

At the same time, the version of the SM with Majorana masses of left - handed neutrinos belongs to the topological class different from that of the SM with Dirac masses of the neutrinos. This means, that the corresponding two systems cannot be connected continuously (at least, if we neglect the CP breaking interactions). For example, the Majorana masses of the left - handed neutrinos may appear as a result of the  type II seesaw \cite{Grimus:2009mm}. Considering the case of the type I seesaw, we come to the conclusion, that the corresponding vacuum is topologically equivalent to the Dirac vacuum without Majorana masses, which follows from the existence of the Majorana masses of the right - handed neutrinos.

\section*{Acknowledgements}

The part of the work of  M.A.Z. performed in Russia  was supported by Russian Science Foundation Grant No 16-12-10059 (Sections  \ref{SymmetricPhase}, \ref{Nambu}, and Appendix) while the part of the work made in France (the remaining part of the text) was supported by Le Studium Institute of Advanced Studies.
G.E.V acknowledges funding from the European Research Council (ERC)
under the European Union's Horizon 2020 research and innovation programme
(Grant Agreement \# 694248).

\section*{Appendix A. Green functions.}

 We consider the two - point Green's function $\cal G$ defined in a certain gauge
corresponding to the gauge fixing condition $O[A] \rightarrow {\rm min}$:
\begin{eqnarray}
{\cal G}^I_J(x,y) & = & \int D\bar{\psi}D\psi D A \exp\{ - S[A,\psi]\} \nonumber
\\ && \exp\{ - \lambda O[A] + {\rm log} \Delta_{FP}[A]\}\bar{\psi}_J(x)
\psi^I(y),\label{G1}
\end{eqnarray}
Here the integral is over the fermionic fields of the SM while indices $I,J$ enumerate the components of $\psi$. $S[A,\psi[$ is the action of the SM, $A$ is the gauge field, the Faddeev - Popov determinant has the form:
\begin{equation}
\Delta_{FP}^{-1}[A] = \int dg \exp\{ - \lambda O[A^g] \}
\end{equation}
Here $g$ is the gauge transformation and $A^g$ is the  transformed gauge field;
it is implied that $\lambda \rightarrow \infty$ at the end.
In our case the elements $P$ of the gauge group $G$ are unitary matrices. That's why $[{\cal G}, {\rm P}]=0$ means that ${\rm
P}^+ {\cal G} {\rm P}={\cal G}$:
\begin{eqnarray}
[{\rm P}^+]_L^J{\cal G}^I_J(x,y){\rm P}_I^K & = & \int D\bar{\psi}D\psi D A \exp\{ -
S[A,{\rm P}^+\psi]\} \nonumber
\\ && \exp\{ - \lambda O[A] + {\rm log} \Delta_{FP}[A]\}\bar{\psi}_L(x)
\psi^K(y)\nonumber\\& = & \int D\psi D A \exp\{ - S[A,\psi]\} \nonumber
\\ && \exp\{ - \lambda O[{\rm P}^+ A {\rm P}] + {\rm log}
\Delta_{FP}[A]\}\bar{\psi}_L(x) \psi^K(y),\label{GP}
\end{eqnarray}
From the last equation we obtain that $[{\cal G}, {\rm P}]=0$ for $\rm P$ from
the center of $G$  for any given gauge.
At the same time when the functional $O[A]$ is invariant under the global gauge transformations, we also have $[{\cal
G}, {\rm P}]=0$ for any ${\rm P} \in G$. The particular case of such a gauge is the
Landau gauge: $O[A] = \int {\bf Tr} A^2 d^4x$. In this
gauge any ${\rm P} \in  G$ commutes with the Green function. In the following we assume, that this gauge is chosen.

\section*{Appendix B. Representation of the SM in terms of Nambu - Gorkov spinors.}

\label{Nambu}

We adopt the notations proposed in \cite{Z2014}.
Left handed doublets and the right - handed doublets of quarks are denoted by $L_K^{\bf a}$ and $R_K^{\bf a}$, where $\bf a$ is the generation index while $K$ is the color index. The
left handed doublets and the right - handed doublets of leptons are ${\cal L}^{\bf a} $ and ${\cal R}^{\bf a} $ respectively.
It will be useful to identify the lepton of each generation as the fourth component of colored quark. Then $L^{\bf a}_{a,4} = {\cal L}^{\bf a}_a$ and $R^{\bf a}_{a,4} = {\cal R}^{\bf a}_a$. So, later we consider the lepton number as the fourth color in the symmetric expressions. We define the analogue of the Nambu - Gorkov spinor
\begin{equation}
{\bf L}^{{\bf a}A}_{aiU} = \left(\begin{array}{c}L^{{\bf a}A}_{ai}\\ \bar{L}^{{\bf a}B}_{c^{\prime}i}\epsilon_{c^{\prime}a}\epsilon^{BA}\end{array} \right)
, \, {\bf R}^{{\bf a}A}_{aiU} = \left(\begin{array}{c}\bar{R}^{{\bf a}B}_{bi}\epsilon_{ba}\epsilon^{BA}\\ R^{{\bf a}A}_{a,i} \end{array} \right),\nonumber
\end{equation}
where $A$ is the usual spin index, $U$ is the Nambu - Gorkov spin index ($U = 1,2$ and $A = 1,2$),  $i$ is the $SU(4)$ Pati - Salam color index (the lepton number is the fourth color), $\bf a$ is the generation index, and $a,b$ are the $SU(2)_L, SU(2)_R$ indices. Both ${\bf R}^{{\bf a}A}_{aiU}$ and ${\bf L}^{{\bf a}A}_{aiU}$ for the fixed values of $a$, $i$, and $\bf a$ compose the four - component Dirac spinors ${\bf R}^{{\bf a}}_{ai}$ and ${\bf L}^{{\bf a}}_{ai}$.
These spinors for the fixed value of $a$ have $(N_c + 1) \times N_g=12$ components. Both ${\bf R}$ and ${\bf L}$ belong to the fundamental representation of $U((N_c + 1) \times N_g)$, where $N_c = 3$ is the number of colors, $N_g = 3$ is the number of generations. Notice, that ${\bf L}$ and $\bar{\bf L}$ ($\bf R$ and $\bar{\bf R}$) are not independent:
$$\bar{\bf L}^{{\bf a}}_{ai} = \epsilon_{ab} \Bigl( {\bf L}^{{\bf a}}_{bi}\Bigr)^T i \gamma^2 \gamma^5\gamma^0, \,
\bar{\bf R}^{{\bf a}}_{ai} = \epsilon_{ab}\Bigl( {\bf R}^{{\bf a}}_{bi}\Bigr)^T i \gamma^2 \gamma^5\gamma^0$$


Here the gamma - matrices act on the spinor space - time indices and are defined in chiral representation
\begin{equation}
\gamma^0 = \left(\begin{array}{cc} 0 & 1 \\1&0 \end{array}\right), \quad \gamma^i = \left(\begin{array}{cc} 0 & \sigma^i\\- \sigma^i&0 \end{array}\right), \quad i = 1,2,3, \quad \gamma^5 = \left(\begin{array}{cc} 1 & 0\\0&-1 \end{array}\right)
\end{equation}
where $\sigma^i$ is the space - time Pauli matrix.

Next, we  arrange the Dirac spinors ${\bf L}^{\bf a}_{ai}, {\bf R}^{\bf a}_{ai}$ in the $SO(4)$ spinor $\Psi$:
\begin{equation}
\Psi^{\bf a}_i = \left(\begin{array}{c} {\bf L}^{\bf a}_{ai}\\{\bf R}^{\bf a}_{ai}\end{array}\right)\label{PSI}
\end{equation}
We introduce the Euclidean $SO(4)$ gamma - matrices $\Gamma^a$ (in chiral representation). The action of the SM gauge fields $e^{i \theta}\in U(1)_Y \subset SU(2)_R$, $U^{(L)}\in SU(2)_L$, $U^{(R)}  = \left(\begin{array}{cc} e^{i\theta}& 0 \\0 & e^{-i \theta} \end{array} \right)\in SU(2)_R$  and $V = \left(\begin{array}{cc}Q e^{i\theta/3}& 0 \\0 & e^{-i \theta} \end{array} \right)\in SU(4)_{\rm Pati\, Salam} \subset U(12)$ (where $Q\in SU(3)$) on the given Majorana spinor is:
\begin{eqnarray}
{\Psi}^{\bf a}_{i} &\rightarrow & \Bigl(V_{ij}\frac{1+\Gamma^5\gamma^5}{2} + \bar{V}_{ij}\frac{1-\Gamma^5\gamma^5}{2}\Bigr)  \left( \begin{array}{cc} U^{(L)}_{} & 0 \\ 0 & U^{(R)}_{}\end{array}\right)  {\Psi}^{\bf a}_{j}\nonumber
\end{eqnarray}
Thus $U^{(L)}, U^{(R)}$ realize the representation of $O(4)\simeq SU(2)_L \otimes SU(2)_R$ while $V$ realizes the representation of the subgroup $SU(4)$ of $U(12)$. The action of the element $R \in U(12)$ of the latter group on the spinor $\Psi^{\bf a}_i$ is $\Psi^{\bf a}_i \rightarrow \Bigl(R^{\bf ab}_{ij}\frac{1+\Gamma^5\gamma^5}{2} + \bar{R}^{\bf ab}_{ij}\frac{1-\Gamma^5\gamma^5}{2}\Bigr)  \Psi^{\bf b}_j$.

Again, $\Psi$ and $\bar{\Psi}$ are not independent:
$$\bar{\Psi}^{{\bf a}}_{i} = \Bigl( {\Psi}^{{\bf a}}_{i}\Bigr)^T i \gamma^2 \gamma^5\gamma^0 \Gamma^4 \Gamma^2 \Gamma^5$$

The Gamma - matrices act on the internal $SU(2)_L$ and $SU(2)_R$ spinor indices rather than on the space - time indices, and are defined in chiral representation
\begin{equation}
\Gamma^4 = \left(\begin{array}{cc} 0 & 1 \\1&0 \end{array}\right), \quad \Gamma^i = \left(\begin{array}{cc} 0 & i \tau^i\\-i \tau^i&0 \end{array}\right), \quad i = 1,2,3, \quad \Gamma^5 = \left(\begin{array}{cc} 1 & 0\\0&-1 \end{array}\right)
\end{equation}
Here $\tau^i$ are the basis elements of $SU(2)_{L,R}$ algebra. It is worth mentioning, that {matrix $\Gamma^5$ distinguishes the left - handed spinors from the right - handed spinors}: $\Gamma^5 {\bf L} = {\bf L}$, and $\Gamma^5 {\bf R} = -{\bf R}$.
In Appendix C we list various symmetry operations that act on the spinors in terms of the Nambu - Gorkov spinor $\Psi$.


\label{Partition}

The partition function for the SM fermions in the presence of the SM gauge fields and the SM Higgs boson written in terms of spinor $\Psi$ has the form:
$Z = \int  D {\Psi} e^{iS}$.
The action $S =  S_K + S_{H}$ contains two terms. The first one is the kinetic term
\begin{equation}
S_K = \frac{i}{2}  \int d^4x\Bigl(\bar{\Psi}^{{\bf a}}_{i} \gamma^{\mu} \nabla_{\mu} {\Psi }^{{\bf a}}_{i}  \Bigr)
 \label{I4l_}
\end{equation}
One can check that this term being written in terms of the original SM fermions is reduced to the conventional SM fermion action (without mass term). Here $\nabla_{\mu}$ is the covariant derivative that includes the gauge field of the model.
The term $S_{H}$ contains the interactions with the SM Higgs field. It appears in the form
\begin{equation}
{\bf H} = \sum_{K = 1,2,3,4} {\bf h}_K \Gamma^K\nonumber
\end{equation}
where ${\bf h}^K \in {\cal R}$. Thus, the Higgs boson appears as the four - component real vector that is transformed under the action of $O(4) \simeq SU(2)_L \otimes SU(2)_R$. The following term would give equal masses to all fermions of the SM:
\begin{equation}
S_H =  \frac{1}{2}\int d^4x\Bigl(\bar{\Psi}^{{\bf a}}_{i}\gamma^5 \Gamma^5 {\bf H} {\Psi}^{{\bf a}}_{i}  \Bigr)\label{SI0}
\end{equation}
Using local $SU(2)_L \subset O(4) \approx SU(2)\otimes SU(2)$ transformation ${L}^{{\bf a}A}_{ai} \rightarrow [U_{\bf h}]_{ab} {L}^{{\bf a}A}_{bi}$ we may fix the unitary gauge, in which
\begin{equation}
{\bf H} = H \Gamma^4,  \quad H = v + h \in {\cal R},\label{unitary}
\end{equation}
where $h$ is the real - valued field of the $125$ GeV Higgs boson while $v$ is the condensate.

In order to obtain different masses for the fermions we need to introduce the matrix of the couplings between the Higgs field and the fermions. The corresponding term in the lagrangian may be written easily in terms of the spinor $\Psi$, but we will not need this expression in our present consideration.


Below we represent the symmetries of space - time spinors in terms of the Nambu - Gorkov spinors introduced above.

\begin{enumerate}

\item{CP transformation}

For the usual Dirac spinors $\psi$ the $CP$ transformation is:
$$ \psi(t,\vec{r}) \to i \gamma^2 \bar{\psi}^T(t,-\vec{r})$$
In the left - right components we have:
$$\psi_L(t,\vec{r}) \to i \sigma^2 \bar{\psi}^T_L(t,-\vec{r}) $$
and $$\psi_R(t,\vec{r}) \to - i \sigma^2 \bar{\psi}^T_R(t,-\vec{r}) $$
In terms of the introduced above spinors we have
$${\bf L}^{{\bf a}}_{ai}(t,\vec{r}) \to - \epsilon_{ab} \gamma^0 {\bf L}^{{\bf a}}_{bi}(t,-\vec{r}), \,
{\bf R}^{{\bf a}}_{ai}(t,\vec{r}) \to  \epsilon_{ab} \gamma^0 {\bf R}^{{\bf a}}_{bi}(t,-\vec{r})$$
and
$${\Psi}^{{\bf a}}_{i}(t,\vec{r}) \to \Gamma^4 \Gamma^2 \gamma^0 {\Psi}^{{\bf a}}_{i}(t,-\vec{r})$$

\item{Time reversal transformation}

In the present paper we follow  the definition of the time reversal transformation accepted in  \cite{ref:LL}. The time reversal transformation contains complex conjugation, which transforms the spinor $\psi$ to its conjugate $\gamma^0\bar{\psi}^T = \psi^*$ similar to the C transformation. At the same time the CPT transformation does not contain this conjugation, and being applied to usual spinors is composed of the multiplication by $\gamma^5$ and the inversion of time and space coordinates.
According to the CPT theorem the T - transformation is equal to CP up to the overall inversion and the change of sign of the right - handed fermions. Therefore, the time reversal transformation results in
$${\Psi}^{{\bf a}}_{i}(t,\vec{r}) \to \Gamma^4 \Gamma^2 \Gamma^5 \gamma^0 {\Psi}^{{\bf a}}_{i}(-t,\vec{r})$$

\item{P transformation}

Parity
$${\Psi}^{{\bf a}}_{i}(t,\vec{r}) \to \Gamma^4 \gamma^0 {\Psi}^{{\bf a}}_{i}(t,-\vec{r})$$

\item{Charge conjugation}

Charge conjugation $C = CP \times P$:
$${\Psi}^{{\bf a}}_{i}(t,\vec{r}) \to \Gamma^2 {\Psi}^{{\bf a}}_{i}(t,\vec{r})$$

\item{CT transformation}
$${\Psi}^{{\bf a}}_{i}(t,\vec{r}) \to \Gamma^4 \Gamma^5 \gamma^0 {\Psi}^{{\bf a}}_{i}(-t,\vec{r})$$

\end{enumerate}


\begin{thebibliography}{105}


 \bibitem{NielsenNinomiya1981}
H.B. Nielsen, M. Ninomiya:
Absence of neutrinos on a lattice.  I - Proof by homotopy theory,
Nucl. Phys. B \textbf{185}, 20  (1981);
Absence of neutrinos on a lattice. II - Intuitive homotopy proof,
Nucl. Phys. B \textbf{193}, 173 (1981).

\bibitem{FrogNielBook}
C.D. Froggatt  and  H.B. Nielsen,
 {\it Origin of Symmetry}, World Scientific, Singapore (1991).

\bibitem{GrinevichVolovik1988}
 P.G. Grinevich, G.E. Volovik,
 Topology of gap nodes in superfluid  3He:  $\pi_4$ homotopy group for 3He-B  disclination,
J. Low Temp. Phys. {\bf 72}, 371  (1988).

\bibitem{Horava2005}
P. Ho\v{r}ava, Stability of Fermi surfaces and $K$-theory, Phys. Rev. Lett.
\textbf{95}, 016405 (2005).

\bibitem{HasanKane2010}
M.Z. Hasan and C.L. Kane,
Topological Insulators,
Rev. Mod. Phys. {\bf 82}, 3045--3067 (2010).

\bibitem{Xiao-LiangQi2011}
Xiao-Liang Qi and Shou-Cheng Zhang,
Topological insulators and superconductors,
Rev. Mod. Phys. {\bf 83}, 1057--1110 (2011).

\bibitem{Schnyder2008}
A.P. Schnyder, S. Ryu, A. Furusaki and A.W.W. Ludwig,
Classification of topological insulators and superconductors in three spatial dimensions,
Phys. Rev. {\bf B~ 78}, 195125 (2008); A.P. Schnyder, S. Ryu, A. Furusaki and A.W.W. Ludwig,
Classification of topological insulators and superconductors,
 AIP Conf. Proc. {\bf 1134}, 10 (2009);
 arXiv:0905.2029.

\bibitem{Kitaev2009}
A. Kitaev,
Periodic table for topological insulators and superconductors,
AIP Conference Proceedings, Volume {\bf 1134}, pp. 22--30 (2009);
  arXiv:0901.2686.


\bibitem{Adler1969}
S. Adler,
Axial-vector vertex in spinor electrodynamics,
Phys. Rev. {\bf 177}, 2426--2438 (1969).

\bibitem{BellJackiw1969}
J.S. Bell   and R. Jackiw,
A PCAC puzzle: $\pi_0\rightarrow\gamma\gamma$ in the $\sigma$ model,
Nuovo Cim. A {\bf 60},  47--61 (1969).

\bibitem{Adler2005}
S.L. Adler,
Anomalies to all orders,
in: {\it Fifty years of Yang-Mills theory},  ed. G. 't Hooft, World Scientific,
187--228 (2005).

\bibitem{BevanNature1997}
 T.D.C. Bevan, A.J. Manninen, J.B. Cook, J.R. Hook et al.,
Momentogenesis by $^3$He vortices: an experimental  analog of primordial baryogenesis,
Nature,  {\bf 386},  689--692 (1997).


\bibitem{Volovik2003}
G.E. Volovik,
{\it The Universe in a Helium Droplet},
Clarendon Press,  Oxford (2003).

\bibitem{AbrikosovBeneslavskii1971}
A.A. Abrikosov and  S.D. Beneslavskii,
 Possible existence of substances intermediate between metals and dielectrics,
 ZhETF  {\bf 59}, 1280--1298 (1970),
JETP {\bf 32},  699 (1971).



\bibitem{Soluyanov2015}
A.A. Soluyanov, D. Gresch, Zhijun Wang, QuanSheng Wu, M. Troyer, Xi Dai, B.A. Bernevig,
Type-II Weyl Semimetals,
 Nature {\bf 527}, 495--498 (2015).

\bibitem{SchnyderBrydon2015}
A.P. Schnyder, P.M.R. Brydon,
Topological surface states in nodal superconductors,
J. Phys.: Condens. Matter {\bf 27}, 243201 (2015).

\bibitem{Mizushima2016}
T. Mizushima, Ya. Tsutsumi, T. Kawakami, M. Sato, M. Ichioka, K. Machida,
Symmetry protected topological superfluids and superconductors - From the basics to $^3$He,
J. Phys. Soc. Jpn. {\bf 85}, 022001 (2016).

\bibitem{Bansil2016}
A. Bansil, Hsin Lin, Tanmoy Das,
Colloquium:  Topological band theory,
Rev. Mod. Phys.{\bf 88}, 021004 (2016).

\bibitem{Yonezawa2016}
Shingo Yonezawa,
Bulk topological superconductors,
 AAPPS Bulletin, {\bf 26},  3 (2016).

\bibitem{ZubkovVolovik2013}
G.E. Volovik and M.A. Zubkov,
On symmetry and topological origin of Weyl particles,
arXiv:1312.1267.


\bibitem{ZubkovVolovik2014a}
G.E. Volovik and M.A. Zubkov,
Emergent Weyl spinors in multi-fermion systems,
Nuclear Physics B {\bf 881}, 514--538  (2014),
arXiv:1402.5700.

\bibitem{ZubkovVolovik2014b}
G.E. Volovik and M.A. Zubkov,
Emergent Weyl fermions and the origin of $i=\sqrt{-1}$ in quantum mechanics,
Pis'ma ZhETF {\bf 99},  552--557 (2014);  JETP Lett. {\bf 99},  481--486  (2014),
arXiv:1404.4084.


\bibitem{Zubkov:2006zb}
  M.~A.~Zubkov,
  The Observability of Z(6) symmetry in the standard model,
  Phys.\ Lett.\ B {\bf 649} (2007) no.1,  91
   Erratum: [Phys.\ Lett.\ B {\bf 655} (2007) no.5-6,  309]
  doi:10.1016/j.physletb.2007.09.026, 10.1016/j.physletb.2007.04.005
  [hep-ph/0609029].


\bibitem{Volovik2009}
 G.E. Volovik,
Fermion zero modes at the boundary of superfluid $^3$He-B,
 Pis'ma ZhETF {\bf 90}, 440--442 (2009); JETP Lett. {\bf 90}, 398--401 (2009);
arXiv:0907.5389;
Topological invariant  for superfluid  $^3$He-B and quantum phase transitions,
Pis'ma ZhETF {\bf 90}, 639--643 (2009);  JETP Lett. {\bf 90}, 587--591 (2009);
arXiv:0909.3084;
"Topological superfluid $^3$He-B in magnetic field and Ising variable",
 Pis'ma ZhETF {\bf 91},  215--219 (2010);  JETP Lett. {\bf 91}, 201--205 (2010);
arXiv:1001.1514.




  \bibitem{Bakker:2003gg}
  B.~L.~G.~Bakker, A.~I.~Veselov and M.~A.~Zubkov,
  Internal structure of discretized Weinberg-Salam model,
  Phys.\ Lett.\ B {\bf 583} (2004) 379
  doi:10.1016/j.physletb.2003.12.062
  [hep-lat/0301011].

  \bibitem{Volovik2000}
  G.E. Volovik,
  Momentum-space topology of Standard Model,
  J. Low Temp. Phys.,  {\bf 119}, 241 -- 247 (2000); hep-ph/9907456.

  \bibitem{Volovik2010}
G.E. Volovik,
 Topological invariants  for Standard Model: from semi-metal to topological insulator,
 Pis'ma ZhETF {\bf 91}, 61--67 (2010);   JETP Lett. {\bf 91}, 55--61 (2010);
arXiv:0912.0502.


 \bibitem{WangYang2016}
Wang Yang, Yi Li,  and Congjun Wu,
Topological septet pairing with spin-3/2
fermions: High-partial-wave channel
counterpart of the $^3$He-B phase,
Phys. Rev. Lett. {\bf 117}, 075301 (2016)




\bibitem{WeinbergQTF}
S. Weinberg,
{\it The Quantum Theory of Fields},
Cambridge University Press, 1996.


\bibitem{Metlitski2014}
M.A. Metlitski, L. Fidkowski, Xie Chen, and A. Vishwanath,
Interaction effects on 3D topological superconductors: surface
topological order from vortex condensation, the 16 fold way and
fermionic Kramers doublets,
arXiv:1406.3032.

\bibitem{Ryu2015}
Shinsei Ryu,
Interacting topological phases and quantum anomalies,
Phys. Scr. T {\bf 164}, 014009  (2015).



\bibitem{Tachikawa2016}
Yuji Tachikawa, Kazuya Yonekura
 Gauge interactions and topological phases of matter,
arXiv:1604.06184.

\bibitem{Witten2015}
E. Witten,
Fermion Path Integrals And Topological Phases,
arXiv:1508.04715.

\bibitem{Yonekura2016}
Kazuya Yonekura,
Dai-Freed theorem and topological phases of matter,
arXiv:1607.01873.

\bibitem{Ryu2016}
Chang-Tse Hsieh, Gil Young Cho, and Shinsei Ryu
Global anomalies on the surface of fermionic symmetry-protected topological phases
in (3+1) dimensions,
Phys. Rev. B {\bf 93}, 075135 (2016).



\bibitem{Wang2014}
Chong Wang, T. Senthil,
Interacting fermionic topological insulators/superconductors in three dimensions,
Phys. Rev. B {\bf 89}, 195124 (2014).


\bibitem{Kapustin2015}
A. Kapustin, R. Thorngren, A. Turzillo, Zitao Wang,
Fermionic symmetry protected topological phases and cobordisms,
JHEP 1512:052,2015, arXiv:1406.7329



\bibitem{Kitaev}
A. Kitaev,
Homotopy-theoretic approach to SPT phases in action: $Z_{16}$ classification of three-dimensional superconductors,
http://www.ipam.ucla.edu/abstract/?tid=12389\&pcode=STQ2015



\bibitem{Z2014}
  M.~A.~Zubkov,
  Modified model of top quark condensation,
  Phys.\ Rev.\ D {\bf 90}, 057501 (2014)
  [arXiv:1405.4067 [hep-ph]].



\bibitem{ref:LL}
V.B.Berestetskii, E.M. Lifshitz, L.P.Pitaevskii {\sl ``Quantum Electrodynimics, Second Edition: Volume 4 (Course of Theoretical Physics)''},
Pergamon Press, Oxford (1982).

\bibitem{Hernandez:2010mi}
  P.~Hernandez,
  ``Neutrino physics,''
  CERN Yellow Report CERN-2010-001, 229-278
  [arXiv:1010.4131 [hep-ph]].



\bibitem{So1985}
H. So,
 Induced topological invariants by lattice fermions in odd dimensions,
Prog. Theor. Phys. {\bf 74}, 585--593 (1985).

\bibitem{IshikawaMatsuyama1986}
K. Ishikawa  and T. Matsuyama,
Magnetic field induced multi component QED in
three-dimensions and quantum Hall effect,
 Z. Phys. C {\bf 33}, 41--45 (1986).

\bibitem{Kaplan1992}
 D.B. Kaplan,
Method for simulating chiral fermions on the lattice, Phys. Lett.  B {\bf 288},
342--347 (1992); arXiv:hep-lat/9206013.

\bibitem{Golterman1993}
 M.F.L. Golterman, K.  Jansen and D.B. Kaplan,
Chern-Simons  currents and chiral  fermions on the lattice,
 Phys.Lett. B {\bf 301}, 219--223 (1993):
arXiv: hep-lat/9209003.


\bibitem{Creutz2008}
M. Creutz, Four-dimensional graphene and chiral fermions,
JHEP 04 (2008) 017;
arXiv:0712.1201.

\bibitem{Kaplan2011}
D.B. Kaplan and Sichun Sun,
Spacetime as a topological insulator: Mechanism for the origin of the fermion generations,
Phys. Rev. Lett. {\bf 108}, 181807 (2012).

\bibitem{Creutz2011}
M. Creutz,
Confinement, chiral symmetry, and the lattice,
Acta Physica Slovaca {\bf 61},  1--127 (2011),
arXiv:1103.3304.

\bibitem{Grimus:2009mm}
  W.~Grimus, L.~Lavoura and B.~Radovcic,
  Type II seesaw mechanism for Higgs doublets and the scale of new physics,
  Phys.\ Lett.\ B {\bf 674}, 117  (2009),
  arXiv:0902.2325.

\bibitem{ZubkovVolovik2012}
M.A. Zubkov, G.E. Volovik,
Momentum space topological invariants for the 4D relativistic vacua with mass gap,
 Nucl. Phys. B {\bf 860}, 295--309  (2012) .

\bibitem{Zubkov2012}
M.A.Zubkov,
Momentum space topology in the lattice gauge theory,
 arXiv:1210.1989.

\bibitem{CFL0}
M. Alford, K. Rajagopal, F. Wilczek,
Color-flavor locking and chiral symmetry breaking in high density QCD,
Nuclear Physics B {\bf 537}, 443--458 (1999).

\bibitem{CFL}
Michael Buballa,
NJL-model analysis of dense quark matter,
Phys. Rept. {\bf 407},  205--376 (2005).

\bibitem{m1}
R. Anglani, M. Mannarelli, M. Ruggieri,
Collective modes in the color flavor-locked phase,
New J. Phys.{\bf 13}, 055002 (2011).

\bibitem{m1m8}
M. Eto, M. Nitta, and N. Yamamoto,
Instabilities of non-Abelian vortices in dense QCD,
Phys. Rev. Lett. {\bf 104}, 161601 (2010);
M. Eto and M. Nitta,
Color magnetic flux tubes in dense QCD,
Phys. Rev. D {\bf 80}, 125007 (2009)

\bibitem{NJLQCD}
M.K. Volkov and A. Radzhabov,
The Nambu–Jona-Lasinio model and its development,
Phys. Usp. {\bf 49}, 551--561 (2006).

\bibitem{EssinGurarie2011}
A.M. Essin and V. Gurarie,
Bulk-boundary correspondence of topological insulators from their Green's functions,
Phys. Rev. B {\bf 84}, 125132 (2011).

\bibitem{Z2016}
M.~A.~Zubkov,
 Wigner transformation, momentum space topology, and anomalous transport,
  Annals Phys.\  {\bf 373} (2016) 298
  doi:10.1016/j.aop.2016.07.011
  [arXiv:1603.03665 [cond-mat.mes-hall]].

\bibitem{Oliveira:2016muq}
  O.~Oliveira, A.~Kızılersu, P.~J.~Silva, J.~I.~Skullerud, A.~Sternbeck and A.~G.~Williams,
  Lattice Landau gauge quark propagator and the quark-gluon vertex,
  arXiv:1605.09632 [hep-lat].

\bibitem{Ballac}
"Thermal Field Theory" (Cambridge Monographs on Mathematical Physics)
by Michel Le Bellac, Cambridge University Press, 1996.

\bibitem{Ghiglieri:2016xye}
  J.~Ghiglieri and M.~Laine,
  Neutrino dynamics below the electroweak crossover,
  JCAP {\bf 1607} (2016) no.07,  015
  doi:10.1088/1475-7516/2016/07/015
  [arXiv:1605.07720 [hep-ph]].


\end{thebibliography}
\end{document}